\documentclass[a4]{epl2}
\usepackage{graphicx}
\usepackage{epsfig}
\usepackage{color}
\usepackage{amssymb}

\title{\textcolor[rgb]{0.00,0.00,1.00}{{Do we need to modify Maxwell's equations?}}}
\shorttitle{\textcolor[rgb]{0.00,0.00,1.00}{Do we need to modify Maxwell's equations?}}

\author{\textcolor[rgb]{0.00,0.00,0.00}{A. I. Arbab}\inst{}\footnote{\textcolor[rgb]{0.00,0.00,0.00}{arbab.ibrahim@gmail.com}}}

\institute{\inst{}
Department of Physics,
College  of Science, Qassim University, P.O. Box 6644, Buraidah 51452, KSA\\
\inst{}
Department of Physics,
Faculty of Science, University of Khartoum, P.O. Box 321, Khartoum 11115, Sudan
}

\abstract{Maxwell's equations are modified to incorporate a scalar field  to account for the London's superconductivity. Assuming the electromagnetic field is described by the Klein-Gordon equation, London's equations of superconductivity are then derived, which are invariant under a new set of transformations. The invariance of the modified Maxwell's equations under these transformations requires the electromagnetic field and the scalar field to be scale-invariant.  Relying on these transformations, a quantized Josephson-like  current is derived. This current gives rise to a residual magnetic field. The spatial and temporal variations of the scalar field are linked to the electric polarization such that the polarization vector is curl-less.}

\pacs{}{}

\begin{document}
\maketitle
\baselineskip=21pt

\section{\textcolor[rgb]{0.00,0.07,1.00}{Introduction}}
Combining Ampere, Gauss, and Faraday equations, Maxwell modified the Ampere equation to unify the phenomenon of electricity and magnetism. He further proved that light is but an electromagnetic field. This great edifice stood firm against all tests that a theory can be confronted with. It is  successful in virtually explaining almost all electromagnetic phenomena. However, London in an attempt to explain the phenomenon of superconductivity had used Maxwell's equations in addition to Newton's law assuming the presence of two non-interacting electric fluids \textcolor[rgb]{0.00,0.07,1.00}{\cite{london}}. He then obtains two fundamental equations known today as London's equation of superconductivity. These equations are  classical, however. It is found by Meissner that the magnetic field can't penetrate the superconductor \textcolor[rgb]{0.00,0.07,1.00}{\cite{messiner}}. It can enter a very small distance known as  London's penetrating depth. The full theory of superconductivity is presented by Bardeen, Cooper and Schrieffer \textcolor[rgb]{0.00,0.07,1.00}{\cite{bcs}}.

London assumed that superconductivity is manifested by the presence of super-electrons besides the ordinary electrons. It is then found that the electromagnetic interaction is of a short-range unlike that one in a vacuum where its range is infinite.
Thus, one can assume that the electromagnetic field is governed by the Klein-Gordon equation, however. Does the question then arise whether it is possible to utterly derive  London's equations from Maxwell's equations without invoking  Newton's second law of motion? The answer is yes if we modify Maxwell's equations without destroying their apparent mathematical beauty. This is what we are about in this work.

To this aim, a new mathematical construct, called the \emph{quaternions}, is employed. To modify Maxwell's equations, we relax the Lorenz gauge condition. We relate it to a scalar function, $\Lambda(r,t)$. The set of equations is found to incorporate this scalar function (field). The spatial and temporal variations of the scalar field give rise to an effect that is similar to the electric polarization of a medium in which the electromagnetic field propagates. This scalar field shows up in the energy and momentum conservation equations. The modified Maxwell's equations state that any temporal or spatial variation of the background (medium) in which the electromagnetic field, an effective electric charge, and current densities arise that change the energy and momentum densities of the electromagnetic field.

Of these backgrounds, are the temperature, gravitational field, phase angle. We can restore the ordinary Maxwell's equations by setting the scalar field to zero. This scalar field introduces an additional force and power to the  charges and current present. We derive  London's equations from the modified Maxwell's equations without resort to  Newton's second law of motion.

The modified Maxwell's equations permit us to allow the electromagnetic field to be described by the Klein-Gordon equation allowing the photon to be massive. This theory is remarkable since it doesn't  destroy the gauge invariance. It  introduces a new gauge-like transformation for the charge and current densities.

Because of the invariance of London's equation under the proposed current transformation, a quantized  current is produced that is  of a Josephson-type current \textcolor[rgb]{0.00,0.07,1.00}{\cite{josephson}}. This is so since in quantum mechanics the current is related to the gradient of the wavefunction that incorporates a phase angle. In Josephson explanation, it is that phase difference that gives rise to the Josephson current \textcolor[rgb]{0.00,0.07,1.00}{\cite{josephson}}. That current is shown not to flow along the electric field direction thus reflecting a tensorial behavior of the electric conductivity of superconductors.

\section{\textcolor[rgb]{0.00,0.07,1.00}{Modified Maxwell's electrodynamics}}

Modified Maxwell's equations are given by \textcolor[rgb]{0.00,0.07,1.00}{\cite{extended, vlaendern,spinorarbab}}
\begin{equation}\label{1}
\vec{\nabla}\cdot\vec{E}-\frac{\rho}{\varepsilon_0}=\frac{\partial\Lambda}{\partial t}\,,\qquad\qquad \qquad\qquad\vec{\nabla}\cdot\vec{B}=0\,,
\end{equation}
and
\begin{equation}\label{1}
\frac{\partial \vec{B}}{\partial t}+\vec{\nabla}\times\vec{E}=0\,,\qquad\qquad \vec{\nabla}\times\vec{B}-\frac{1}{c^2}\,\frac{\partial\vec{E}}{\partial t}=\mu_0\vec{J}-\vec{\nabla}\Lambda\,,
\end{equation}
where
\begin{equation}
-\Lambda=\vec{\nabla}\cdot\vec{A}+\frac{1}{c^2}\frac{\partial\varphi}{\partial t}\,,
\end{equation}
is the modified Lorenz gauge condition. The above equations reduce to the ordinary Maxwell's equations when the Lorenz gauge condition is satisfied, \emph{i.e.}, $\Lambda=0$.

The energy conservation equation can be derived from the modified Maxwell's equations above which reads
\begin{equation}\label{1}
\frac{\partial u_\Lambda}{\partial t}+\vec{\nabla}\cdot\vec{S}_\Lambda=-\vec{E}\cdot\vec{J}-c^2\rho\Lambda\,,\qquad\qquad \vec{S}_\Lambda=\frac{\vec{E}\times\vec{B}-\Lambda\vec{E}}{\mu_0}\,,\qquad u_\Lambda=\frac{\varepsilon_0}{2}E^2+\frac{B^2}{2\mu_0}+\frac{\Lambda^2}{2\mu_0}\,.
\end{equation}
Equation (4) reveals that an extra energy $(\frac{\Lambda^2}{2\mu_0}$) and flux ($-\mu_0^{-1}\Lambda\, \vec{E}$) are associated with the $\Lambda$ field. This extra field is thus coupled to the electromagnetic field. It satisfies the wave equation
\begin{equation}
\frac{1}{c^2}\frac{\partial^2\Lambda}{\partial t^2}-\nabla^2\Lambda=0\,.
\end{equation}
The momentum conservation equation can be derived from the modified Maxwell's equations above which reads
\begin{equation}\label{1}
-\left(\frac{\partial \vec{g}_\Lambda}{\partial t}\right)_i=\frac{\partial \sigma^\Lambda_{ij}}{\partial x_j}+(\rho\,\vec{E}+\vec{J}\times\vec{B}+\Lambda\,\vec{J})_i\,.
\end{equation}
and
\begin{equation}\label{1}
\vec{g}_\Lambda=\varepsilon_0(\vec{E}\times\vec{B}+\Lambda\vec{E})\,\,,\qquad \sigma^\Lambda_{ij}=\left(\frac{\varepsilon_0E^2}{2}+\frac{B^2}{2\mu_0}-\frac{\Lambda^2}{2\mu_0}\right)\delta_{ij}-\varepsilon_0E_iE_j-\mu_0^{-1}(B_iB_j+\epsilon_{ijk}\Lambda B_k)\,.
\end{equation}
Here $\vec{g}$ and $\sigma^\Lambda_{ij}$ are the momentum density and stress tensor of the electromagnetic field which is coupled to the field $\Lambda$. The modified Lorentz force density is
\begin{equation}
\vec{f}_\Lambda=\rho\vec{E}+\vec{J}\times\vec{B}+\Lambda\vec{J}\,.
\end{equation}
While the Maxwellian stress tensor is symmetric, the above stress tensor ($\sigma^\Lambda_{ij}$) is not. Moreover, the linear relationship between the momentum density ($\vec{g}$) and the Poynting vector ($\vec{S}$), \emph{viz}.,  $\vec{g}=\frac{\vec{S}}{c^2}$, is not valid, however.

The modified Lorenz force and power of a moving charge $q$ are accordingly given by \textcolor[rgb]{0.00,0.07,1.00}{\cite{extended,vlaendern}}
\begin{equation}\label{1}
\vec{F}_\Lambda=q\vec{E}+q\vec{v}\times\vec{B}+\Lambda\,q\vec{v}\,\,,\qquad\qquad\qquad P_\Lambda=q\vec{v}\cdot\vec{E}+qc^2\Lambda\,.
\end{equation}
The last term in the above force, \emph{viz}., $\Lambda q\vec{v}$ is of a viscous-like force that dictates the present of a fluid permeating the space. This fluid (\emph{ether}) had once been introduced to allow the electromagnetic wave to propagate through it, but later rejected. One can therefore, treat $\Lambda$ as  a background in which the electromagnetic field propagates. It influences the electromagnetic field when this field is not uniform and homogenous.

Examples of this background could be the temperature and the gravitational field. It is well known that the phenomenon of thermoelectricity is associated with a temperature gradient. We here generalized the thermal effects that could occur before the  electromagnetic system comes to thermal equilibrium. We argue that Eqs.(1) and (2) could be  Maxwell's equations in a non-isothermal medium. Such a scheme could lead to a theory thermoelectricity \textcolor[rgb]{0.00,0.07,1.00}{\cite{thermo}}. At a microscopic level, a non-uniform phase angle can induce observable electromagnetic effects. A temporal variation of the background quantity represented by $\Lambda$ leads to an effective charge density ($\varepsilon_0\frac{\partial\Lambda}{\partial t}$), while the spatial variation leads to an effective current, ($\mu_0^{-1}\vec{\nabla}\Lambda$). Thus, these variations will lead to significant charge and current densities if were done during short times and distances.

At the same time, if $\Lambda$ refers to a non-uniform gravitational field, then such a non-uniformity will be imprinted in the electromagnetic field. Consequently, the electromagnetic field propagating in a non-uniform gravitational field will be influenced. Einstein's theory of general relativity had shown that light propagating in a non-uniform gravitational field undergoes a frequency shift. Einstein attributed that to the curvature of space-time.

\section{\textcolor[rgb]{0.00,0.07,1.00}{Klein-Gordon equation and London's equations}}

The scalar field $\Lambda$ could also satisfy the Klein-Gordon equation,
\begin{equation}\label{1}
\frac{1}{c^2}\frac{\partial^2\Lambda}{\partial t^2}-\nabla^2\Lambda+\left(\frac{mc}{\hbar}\right)^2\Lambda=0\,,
\end{equation}
if we let
\begin{equation}\label{1}
\vec{\nabla}\cdot\vec{J}+\frac{\partial\rho}{\partial t}=\frac{\Lambda}{\mu_0}\left(\frac{mc}{\hbar}\right)^2\,.
\end{equation}
With some scrutiny, and since the scalar field is also coupled to the current and charge, as evident from Eqs.(4) and (6), we will see that the total charge of the system  is conserved, if we employ Eq.(3). Thus,
\begin{equation}\label{1}
\vec{\nabla}\cdot\vec{J}_T+\frac{\partial\rho_T}{\partial t}=0\,,
\end{equation}
where
\begin{equation}\label{1}
\vec{J}_T=\vec{J}+\mu_0^{-1}\beta^2\vec{A}\,,\qquad\qquad \rho_T=\rho+\varepsilon_0\beta^2\varphi\,,\qquad \beta=\frac{mc}{\hbar}\,.
\end{equation}
It is interesting to see the right equation expressing the conservation of charge is now Eq.(12).

There are some instances where the electric and magnetic fields satisfy the Klein-Gordon equation. This occurs inside the superconductor, where the electromagnetic interaction becomes of short range that may dictate that photons become massive. One can entertain this opportunity and derive the wave equations
\begin{equation}\label{1}
\frac{1}{c^2}\frac{\partial^2\vec{E}}{\partial t^2}-\nabla^2\vec{E}=-\mu_0\left(\frac{\partial\vec{J}}{\partial t}+\vec{\nabla}\rho c^2\right),
\end{equation}
and
\begin{equation}\label{1}
\frac{1}{c^2}\frac{\partial^2\vec{B}}{\partial t^2}-\nabla^2\vec{B}=\mu_0\vec{\nabla}\times\vec{J}\,,
\end{equation}
\begin{equation}\label{1}
\frac{1}{c^2}\frac{\partial^2\Lambda}{\partial t^2}-\nabla^2\Lambda=-\mu_0\left(\frac{\partial\rho}{\partial t}+\vec{\nabla}\cdot\vec{J}\right).
\end{equation}
employing Eqs.(1) and (2). If the electromagnetic field had to satisfy the Klein-Gordon equation, then Eq.(14), (15) and (16) should read
\begin{equation}\label{1}
\mu_0\left(\frac{\partial\vec{J}}{\partial t}+\vec{\nabla}\rho c^2\right)=\left(\frac{mc}{\hbar}\right)^2\vec{E}\,,
\end{equation}
\begin{equation}\label{1}
\mu_0\vec{\nabla}\times\vec{J}=-\left(\frac{mc}{\hbar}\right)^2\vec{B}\,,
\end{equation}
and
\begin{equation}\label{1}
\mu_0\left(\frac{\partial\rho}{\partial t}+\vec{\nabla}\cdot\vec{J}\right)=\left(\frac{mc}{\hbar}\right)^2\Lambda\,.
\end{equation}
Notice that Eq.(19) is shown above to comply with the charge conservation. Solving Eqs.(17) - (19) shows that $\rho$, $\vec{J}$, $\Lambda$, $\vec{E}$ and $\vec{B}$ satisfy the Klein-Gordon equation. They exhibit the Meissner's effect when they are stationary.  It is remarkable that Eqs.(17) and (18) are the London's equations for superconductivity. They were derived by London's brothers using different physical principles. London used Newton's second law of motion and an idea of super-electrons, having a number density,  $n_s$. Equations (17) and (18) are gauge invariant. It is remarkable that Eqs.(17) - (19) lead to Meissner effect of superconductivity, \emph{i.e.}, the decay of the static electromagnetic field with distance inside the superconductor, \emph{i.e}., the magnetic field doesn't penetrate the superconductor. We have here the London penetration depth, $\lambda_L=\frac{\hbar}{mc}$, whereas in London's theory, $\lambda_L=\sqrt{\frac{m_s}{\mu_0 n_se_s^2}}$\,, where $m_s$ is the mass of the super-electron and $e_s$ its charge. These two relations set a relation between the electron mass and the photon mass by
\begin{equation}\label{1}
m=e_s\hbar\sqrt{\frac{\mu_0n_s}{m_sc^2}}\,.
\end{equation}
Using the analogy between matter wave and electromagnetic wave, we  have found recently that the electric conductivity is related to the photon mass by the relation, $\sigma=\frac{2m}{\mu_0\hbar}$ \textcolor[rgb]{0.00,0.07,1.00}{\cite{analogy}}. Hence, upon using Eq.(20), we arrive at the interesting relation
\begin{equation}\label{1}
\sigma=2e_s\sqrt{\frac{\epsilon_0n_s}{m_s}}\,.
\end{equation}
If super-electron superconductivity is that of the electron in a conductor, then one may write
\begin{equation}\label{1}
\sigma=\frac{n_se_s^2\tau_s}{m_s}\,,
\end{equation}
where $\tau$ is the relaxation time of super-electrons inside a superconductor. Are the two electric superconductivities equal? If so, one has the relaxation time for super-electrons as
\begin{equation}\label{1}
\tau_s=\sqrt{\frac{4\epsilon_0m_s}{n_se_s^2}}\,,\qquad\qquad \omega_s=\sqrt{\frac{\pi n_se_s^2}{\varepsilon_0m_s}}\,,
\end{equation}
where, $\omega_s=\frac{2\pi}{\tau_s}$, is the plasma frequency of super-electrons.

Let us study the magnetic flux encapsulated by an electric  current. This can be found using Eq.(18), and integrate it over the surface through which the current flows and the magnetic field emanates. Therefore, Eq.(18) yields
\begin{equation}\label{1}
\int\vec{J}\cdot d\vec{\ell}=-\frac{m^2c^2}{\mu_0\hbar^2}\,\phi_B\,,
\end{equation}
where $\phi_B$ is the magnetic flux. The above minus sign has to do with the Lenz's rule. If the flux is now quantized, \emph{i.e}., $\phi_B=n\frac{h}{e}$, where $n=1, 2, 3, \cdots$, then Eq.(24) yields
\begin{equation}\label{1}
\int\vec{J}\cdot d\vec{\ell}=-\,\frac{m^2c^2}{\mu_0e\hbar}\,2\pi n\,,
\end{equation}
where closed loops imply that $m=0$, otherwise loops are not closed.

 Recall that Eqs.(17) - (19) are invariant under the transformations
\begin{equation}\label{1}
\vec{J}\,'=\vec{J}+\vec{\nabla}\chi\,,\qquad\qquad \rho\,'=\rho-\frac{1}{c^2}\frac{\partial\chi}{\partial t}\,,
\end{equation}
where $\chi$ is some scalar function. Invariance of  Eqs.(1) and (2) under the transformation in Eq.(26),  yields $\Lambda=\mu_0\chi$. The transformations in Eq.(26) are analogous to the gauge transformations
\begin{equation}\label{1}
\vec{A}\,'=\vec{A}+\vec{\nabla}f\,,\qquad\qquad \varphi\,'=\varphi-\frac{\partial f}{\partial t}\,,
\end{equation}
for some scalar function, $f$.
Therefore, even $\int \vec{J}\cdot d\vec{\ell}=0$, but $\int \vec{\nabla}\chi\cdot d\vec{\ell}\ne 0$. Therefore, Eq.(25) would imply that
\begin{equation}\label{1}
\int\vec{\nabla}\chi\cdot d\vec{\ell}=\Delta\,\chi=\,\frac{m^2c^2}{\mu_0e_s\hbar}\,2\pi n\,,
\end{equation}
which when Eq.(20) is used, becomes
\begin{equation}\label{1}
\Delta\,\chi=\frac{n_s\,e_sh}{m_s}\,h\, n\,.
\end{equation}
In terms of $\Lambda$, Eq.(29) can be expressed as
$$\hspace{7cm}\Delta\,\Lambda=\frac{\mu_0e_sn_s}{m_s}\, h\,n\,, \hspace{6cm}(A)$$
where $\Lambda$ has a dimension of magnetic field (magnetic field scalar).
In general,  Eq.(24) implies that
\begin{equation}\label{1}
\Delta \chi=\frac{m^2c^2}{\mu_0\hbar^2}\,\phi_B\,.
\end{equation}
Note that in quantum mechanics, the function $\chi$ appears in the phase of the wavefunction representing the quantum particle. The current in quantum mechanics is defined as a gradient of the wavefunction that involves the phase of the wavefunction. This shows that the phase of a wavefunction is a measurable quantity. Such kind of a phase appears in Josephson effect \textcolor[rgb]{0.00,0.07,1.00}{\cite{josephson}}.
Thus, a quantized  phase current, $J_p=\nabla\chi$,  will be \footnote{Eq.(18) suggests a current depending on the magnetic field of the form, $J_p=\frac{n_se_s^2B\Delta x}{m_s}$.}
\begin{equation}\label{1}
J_p\equiv\frac{\Delta \chi}{\Delta x}=\left(\frac{e_sn_sh}{\Delta x \,m_s}\right)n\,.
\end{equation}
Moreover, Eq.(A) could be associated with some  minimum (critical) magnetic field so that $B_\Lambda=\frac{\mu_0n_se_sh}{m_s}$, with an intensity of $H_\Lambda=\frac{n_se_sh}{m_s}$\,. It can be related to the London's penetration depth by the relation $B_\Lambda=\frac{h}{e_s}\,\frac{1}{\lambda^2_L}$.
The quantum current in Eq.(31) can be expressed as
$$J_p=\frac{h}{\mu_0e_s\Delta x}\,\frac{n}{\lambda_L^2}\,.$$
It is therefore evident  that the quantized magnetic flux  gives rise to a quantized current. In Josephson effect, $\Delta x$, can be considered as the width of the junction, a distance over which the phase changes. In the standard Josephson effect, super-electron (Cooper pairs) tunnel quantum mechanically through the insulating payer making the junction (superconductor-insular-superconductor). In fact, this is not the only possible explanation, the field-theoretic interpretation, where massive photons propagate through the junction, also yields similar results.

Let us now express the electric and magnetic fields, in terms of the scalar and vector potentials, $\varphi$ and $\vec{A}$, as
 \begin{equation}\label{1}
\vec{E}=-\frac{\partial\vec{A}}{\partial t}-\vec{\nabla}\varphi\,,\qquad\qquad \vec{B}=\vec{\nabla}\times\vec{A}\,.
\end{equation}
 Using Eq.(32),  Eqs.(17) and (18) yield
 \begin{equation}\label{1}
\vec{J}=-\frac{m^2c^2}{\mu_0\hbar^2}\, \vec{A}\,,\qquad\qquad \rho=-\frac{m^2}{\mu_0\hbar^2}\,\varphi\,.
\end{equation}
It is interesting that Eq.(33) is consistent with the transformations in Eq.(13) with $\vec{J}_T=0$ and $\rho_T=0$. Hence, the Lorenz gauge condition and the conservation of charge (continuity equation) are different manifestation of the same entity. It is interesting to see that Eq.(3) is invariant under the transformations in Eq.(27). Now apply Eq.(26) in Eq.(8) to see whether the Lorentz force density is invariant under these transformations or not. This yields
\begin{equation}
\vec{f}\,'_\Lambda=\vec{f}_\Lambda+\left(\vec{\nabla}\times (\chi \vec{B})-\frac{1}{c^2}\frac{\partial(\chi\vec{E})}{\partial t}+\vec{\nabla}(\chi \Lambda)-\mu_0(\chi\vec{J})\right)\,.
\end{equation}
If the Lorentz force density is invariant under the transformations (26), then the electric field, magnetic field,  and the scalar field must be invariant under the scale invariance
\begin{equation}
 \vec{B}\rightarrow \chi\vec{B}\,,\qquad \vec{E}\rightarrow \chi\vec{E}\,\,,\qquad \Lambda\rightarrow \chi\Lambda\,.
\end{equation}
Under this condition the Lorentz force, $\vec{f}\,'=\vec{f}_\Lambda$, upon using Eq.(2) in Eq.(34). However, we see that under the general scaling, $\rho'=\chi \rho, \vec{J}'=\chi\vec{J}, \Lambda'=\chi\Lambda, \vec{E}'=\chi\vec{E}\,, \vec{B}'=\chi\vec{B}$, then the force density, as given by Eq.(8), is, $\vec{f}\,'=\chi^2\vec{f}_\Lambda$.

Let us now consider the invariance of Eqs.(1) and (2) under the above transformations. This yields
\begin{equation}
 (\vec{\nabla}\chi)\times\vec{B}+\Lambda\vec{\nabla}\chi-\frac{\vec{E}}{c^2}\frac{\partial\chi}{\partial t}=0\,,\qquad (\vec{\nabla}\chi)\cdot\vec{E}=\Lambda\frac{\partial\chi}{\partial t}\,,\qquad (\vec{\nabla}\chi)\cdot\vec{B}=0\,,\qquad(\vec{\nabla}\chi)\times\vec{E}+\vec{B}\,\frac{\partial\chi}{\partial t}=0\,.
 \end{equation}
A consistency of the above system requires, $\Lambda^2=\frac{E^2}{c^2}-B^2$. This shows that the electric and magnetic fields are no longer orthogonal to the direction of propagation. From Eq.(26) one can define the charge and current densities, respectively, as, $\rho_p=-\frac{1}{c^2}\frac{\partial\chi}{\partial t}$ and $ \vec{J}_p=\vec{\nabla}\chi$ that are determined by Eq.(36) as
\begin{equation}
\vec{J}_p= -\frac{\vec{J}_p\times\vec{B}}{\Lambda}-\frac{\rho_p}{\Lambda}\,\vec{E}\,,\qquad \vec{J}_p\cdot\vec{E}=-c^2\Lambda\rho_p\,,\qquad \vec{J}_p\cdot\vec{B}=0\,,\qquad \vec{J}_p\times\vec{E}=c^2\rho_p\,\vec{B}\,.
 \end{equation}
 In addition to the fact that $\vec{J}_p$ is perpendicular to the magnetic field $\vec{B}$, but not to the electric field, $\vec{E}$, Eq.(37) also reveals that $J_p=c\rho_p$\,. Equation (37) is reminiscent of the Hall effect where when  a magnetic field is applied at right angles to the direction of a current flowing in a conductor, an electric field is created in a direction perpendicular to both.
 The above equation is also found to emerge from treating photons to be massive inside a medium \textcolor[rgb]{0.00,0.07,1.00}{\cite{massivephotons}}. The additional current, $\vec{J}_p$,  flows along a direction that makes an angle different from $\pi/2$ with the electric field. This behavior could be attributed to the fluid prevailing inside the superconductor, which we mentioned before, that deflects the moving super-electrons. It is interesting to note that the electric current in a conductor is always along the electric field direction. This immediately reflects the tensorial behavior of the electric conductivity of the superconductor.

\section{\textcolor[rgb]{0.00,0.00,1.00}{Relation of the electric polarization to the scalar field}, $\Lambda$}

Maxwell's equations inside a material is often expressed as
\begin{equation}\label{1}
\vec{\nabla}\cdot\vec{E}=\frac{\rho}{\varepsilon_0}-\frac{\vec{\nabla}\cdot\vec{P}}{\varepsilon_0}\,,\qquad\qquad \qquad\qquad\vec{\nabla}\cdot\vec{B}=0\,,
\end{equation}
and
\begin{equation}\label{1}
\vec{\nabla}\times\vec{B}=\frac{1}{c^2}\,\frac{\partial\vec{E}}{\partial t}+\mu_0\left(\vec{J}+\frac{\partial\vec{P}}{\partial t}\right)\,,\qquad\qquad  \frac{\partial \vec{B}}{\partial t}+\vec{\nabla}\times\vec{E}=0\,,
\end{equation}
where $\vec{P}$ is the electric polarization vector. Comparing Eqs.(38) and (39) with Eqs.(1) 1nd (2) yields
\begin{equation}\label{1}
-\vec{\nabla}\cdot\vec{P}=\varepsilon_0\frac{\partial\Lambda}{\partial t}\,,\qquad \qquad \mu_0 \frac{\partial\vec{P}}{\partial t}=-\vec{\nabla}\Lambda\,,
\end{equation}
which leads to
\begin{equation}\label{1}
\frac{1}{c^2}\frac{\partial^2\Lambda}{\partial t^2}-\nabla^2\Lambda=0\,.
\end{equation}
We deduce, from Eq.(40), the following equations
\begin{equation}\label{1}
\vec{\nabla}\times\vec{P}=0\,,\qquad \qquad  \frac{1}{c^2}\frac{\partial^2\vec{P}}{\partial t^2}-\nabla^2\vec{P}=0\,.
\end{equation}
Thus, the polarization vector satisfies a wave equation traveling at the velocity of light.

 Let us now use Eq.(40) to derive the conservation equation
\begin{equation}\label{1}
\vec{\nabla}\cdot(c^2\Lambda\vec{P})+\frac{\partial}{\partial t} \left(\frac{P^2}{2\varepsilon_0}+\frac{\Lambda^2}{2\mu_0}\right)=0\,.
\end{equation}
It is interesting to see that the polarization and the scalar field, $\Lambda$ interact and produce a wave pattern in the medium. The polarization energy is transmitted along the polarization vector. The Poynting vector of this wave is $c^2\Lambda\,\vec{P}$. The energy conservation equation associated with Eqs.(38) and (39) is
\begin{equation}\label{1}
\vec{\nabla}\cdot\vec{S}_P+\frac{\partial u_P}{\partial t} =-\vec{J}\cdot\vec{E}_P\,,
\end{equation}
where
\begin{equation}
\vec{S}_P=\frac{\vec{E}_P\times\vec{B}}{\mu_0}\,,\qquad\qquad u_P=\frac{1}{2}\,\varepsilon_0E_P^2+\frac{B^2}{2\mu_0}\,,\qquad \vec{E}_P=\vec{E}+\frac{\vec{P}}{\varepsilon_0}\,,
\end{equation}
where  $\vec{P}$ satisfies Eq.(42). Equations (44) and (45) show that the effect of polarization on the energy equation is to modify $\vec{E}$ by $\vec{E}_P$ to include the polarization contribution. Interestingly, if the polarization vector doesn't point along the electric field direction, then some of the electromagnetic energy will  flow along  it, as well.

Applying Eq.(3) in Eq.(40) yields the two equations
\begin{equation}\label{1}
\vec{\nabla}\cdot\left(q\frac{\partial\vec{A}}{\partial t}-\frac{q\vec{P}}{\varepsilon_0}\right)+\frac{\partial }{\partial t}\left(\frac{\partial }{\partial t}\frac{q\varphi}{c^2}\right)=0\,,\qquad \qquad  \frac{\partial}{\partial t}\left(\vec{\nabla}(q\varphi)-\frac{q\vec{P}}{\varepsilon_0}\right)+\vec{\nabla}(\vec{\nabla}\cdot qc^2\vec{A})=0\,.
\end{equation}
 Equation (46) can be read as representing an energy conservation equation, where the term $\frac{\partial}{\partial t}\frac{q\varphi}{c^2}$ is the mass creation rate followed by the charge creation by polarizing the medium. In fact, while we create charges from the medium (material), we create at the same time masses. Thus, energy conservation should govern this mechanism.

\section{\textcolor[rgb]{0.00,0.07,1.00}{Concluding remarks}}

We have shown in this work how modifying Maxwell's equations without disturbing their current beauty could bring new impetus to electromagnetism. Some fundamental phenomena could be explained using the new electrodynamics. These phenomena are understood to represent quantum macroscopic effects. The way they are interpreted is very intriguing and exciting. In  ordinary electromagnetism, the effect of the background in which the electromagnetic field is present was not taken into consideration. Before, the effect of the medium is embedded in redefining the permittivity and permeability of the medium. As in general relativity, the nature of the space-time in which electromagnetic phenomena occur is of  paramount importance and  should be here too.  The non-uniformity (in space and time) in the medium can be parameterized by the scalar $\Lambda$, \emph{e.g}., non-isothermal equilibrium, non-pressure equilibrium,  quantum non-phase equilibrium. All of these non-uniformities can lead to significant observable effects. Moreover, a straightforward derivation of London's equations is found without resort to Newton's second law in describing electrons dynamics.

\end{document}